\documentclass[aps,prb,twocolumn,floatfix,showpacs]{revtex4}
 \pdfoutput=1

\usepackage{graphicx} 
\usepackage{calc} 
\usepackage{amsfonts}  
\usepackage{amsmath}

\newcommand{\betrag}[1]{\left| #1 \right|}

\begin{document}

\title{  Link between antiferromagnetism and   superconductivity probed by nuclear spin  relaxation in organic conductors}        
 
\author{  C. Bourbonnais$^{1,2}$ and A. Sedeki$^1$}
\affiliation{$^1$Regroupement Qu\'ebecois sur les Mat\'eriaux de Pointe,
  D\'epartement de physique, Universit\'e de Sherbrooke, Sherbrooke,
  Qu\'ebec, Canada, J1K-2R1 }
  \affiliation{$^2$ Canadian Institute for Advanced Research Toronto, Ontario M5G 1Z8, Canada  }

\date{\today}

\begin{abstract}
The interdependence of antiferromagnetism and superconductivity  in the Bechgaard salts series of organic conductors is examined in the light of the  anomalous temperature dependence of the nuclear spin-lattice relaxation rate. We apply the renormalization group approach to the  electron gas model  to show  that   the   crossover  from   antiferromagnetism to   superconductivity along with the anomalous  nuclear  relaxation rate of  the Bechgaard salts        can be   well  described  within a unified microscopic  framework. For sizable nesting deviations of the Fermi surface,  scaling theory  reveals how      pairing   correlations enhance short-range antiferromagnetic correlations {\it via} magnetic Umklapp scattering over  a large part  of the  metallic phase  that precedes  superconductivity.  These enhanced magnetic correlations are responsible for the Curie-Weiss behavior  observed in the NMR relaxation rate.\end{abstract}
\pacs{74.20 Mn, 74.70 Kn, 76.60 Es}
\maketitle 

 \section{Introduction}
 In the  attempt to understand how  itinerant  antiferromagnetism can give rise to   superconductivity in correlated electron systems, one  faces  the  difficulty  of    linking    the behavior of  spin  fluctuations   that  can be  extracted from experiments  to the mechanism of  pairing that leads to  superconductivity. This work is about  the quest  of 
  such a connection in  the Bechgaard salts [(TMTSF)$_2$X] series of       organic  conductors.  We ground our analysis on scaling theory,  which allows   a  reexamination of     the  nuclear spin relaxation  in   the metallic phase of  these low dimensional molecular systems. 
 
The  (TMTSF)$_2$X compounds are      quasi-one-dimensional (quasi-1D) conductors   known to exhibit  a spin-density-wave (SDW)  state adjacent to   superconductivity (SC) in their phase diagram \cite{Jerome82}$^,$\cite{Bourbon08,Brown08}.    This particular sequence of states is achieved   by the application of  hydrostatic    pressure   or   by chemical means from  anion  X  substitution.  Among the   host of experimental tools used   to study this  pattern of phases, the   Nuclear Magnetic Resonance (NMR) technique  takes on particular importance due to  its sensitivity  to spin correlations  \cite{Bourbon84,Creuzet87b,Takigawa86,Wzietek93,Wu05,Shinagawa07}.   
 In the metallic state of  the  SC side of the phase diagram,  NMR  measurements have revealed the existence of an    anomalous enhancement   of  the nuclear spin-lattice relaxation rate  $T_1^{-1} $.    The enhancement was  first  observed for  the ambient pressure superconductor (TMTSF)$_2$ClO$_4$\cite{Bourbon84,Creuzet85,Takahashi84}, and subsequently found by  different  groups   to be     a common characteristic  of the  series  in the metallic state  above  the critical pressure for superconductivity\cite{Creuzet87b,Wzietek93,Wu05,Shinagawa07}.   The $ T_1^{-1}$ temperature profile shows   pronounced deviations from the  Korringa  law, $ T_1^{-1}\propto T $, which is  normally expected in conventional  metals.    These deviations were originally ascribed to  the presence of short-range antiferromagnetic spin fluctuations     extending in temperature    dozens of  times the superconducting
   \hbox{$T_c\sim 1$}K in the metallic state\cite{Bourbon84}. The  amplitude of the deviations are strongly pressure dependent and seemingly  tied to the amplitude   of  $T_c$   \cite{Creuzet85a,Creuzet87b,Brown08},     suggesting   that  antiferromagnetism and Cooper pairing are  closely  related.  
  
It was initially proposed that one-dimensional  short-range antiferromagnetic correlations are a key  determinant in  the   enhancement of $T_1^{-1}$. This low dimensional  response  was shown to ultimately cross  over  into a  higher dimensional metallic phase around 10K, below which  a Korringa law was predicted to be  recovered. \cite{Bourbon84} However, the dimensional aspects of  such an  interpretation  run   into    difficulties  when one compares  the  temperature  scale for the crossover    to the range of values   otherwise 
extracted from     experiments\cite{Jacobsen81,Moser98,Vescoli98}, and   which would rather place this scale an order of magnitude  higher  in temperature.    An additional defect comes  from an important observation made by Brown {\it et al.,} \cite{Brown08} concerning the temperature profile of the relaxation. Based on the analysis of recent   measurements,\cite{Wu05,Shinagawa07, Brown08} it was shown that the Korringa behavior is in effect not recovered down to the lowest temperature preceding superconductivity. Deviations  actually    persist indicating that  staggered spin fluctuations, though non singular, keep growing as the temperature is lowered,    imposing   a Curie-Weiss (C-W)   temperature dependence for the relaxation rate. 
  Noticeably enough, however,  the characteristics of the C-W response persists down to $T_c$ and the amplitude of the anomaly as a whole  evolves rapidly under pressure like  the temperature  scale for the onset of superconductivity. The  question then arises if,   in accordance with the customary view, a sharp distinction can be drawn  between     both phenomena, or if  they are in effect dynamically  linked,  a possibility that would not only connect  magnetism to  superconductivity, but  also involve    superconductive pairing   in  the enhancement of spin correlations. 
 
  It is from  the   latter perspective  that we shall  reconsider  the $T_1^{-1}$  problem for the  Bechgaard salts\cite{Ogazaki}.   This will be  achieved with the help of the  weak coupling renormalization group (RG) theory.\cite{Duprat01,Nickel06}  Recent    developments along these lines   have demonstrated  how the  one-loop RG can   take into account        density-wave and Cooper pairings on equal footing in  correlated quasi-1D metals. In the framework of the repulsive  quasi-1D electron gas model, it was  found that unconventional singlet `d-wave' (SCd) or  in certain conditions  triplet `f-wave' (SCf) superconductivity can be dynamically generated next to a SDW state as   alterations of the nesting of the Fermi surface -- which  mimic  pressure effect --  are made sufficiently large. From a similar approach to be brought forward here, a  C-W  type  behavior  for the spin fluctuation response is shown to take  place over a large temperature interval above   $T_c$ within  a SCd scenario. It originates from   magnetic Umklapp scattering whose amplitude is apparently strengthened by constructive interference with  superconductive  pairing. The amplitude of spin correlations rapidly decline under `pressure' , in line with the decrease of $T_c$.   
  When transposed into a $T_1^{-1}$  calculation, the  RG results  can  give a satisfactory account of    the key  features shown by  the nuclear relaxation rate, establishing  a direct   connexion between spin fluctuations and the mechanism of superconductivity in the Bechgaard salts.
 
 In Sec.~II we introduce the quasi-1D electron gas model in the presence of weak Umklapp scattering and alteration of nesting.  We review   the  results of the three-variables RG method obtained at the one-loop level  within a SDW-SCd scenario.  The temperature profile of the  antiferromagnetic and superconducting  responses are given  and scrutinized as a function of   nesting deviations of the Fermi surface. The temperature scales  are extracted and  used to construct  the phase diagram of the model. In Sec.~III, the explicit form for  $T_1^{-1}$ is calculated from the RG results for both   the  antiferromagnetic and uniform components of spin fluctuations. In Sec.~IV we discuss the results  and conclude. 

\section{Itinerant antiferromagnetism and superconductivity: Renormalization group results }
We consider the electron gas model\cite{Dzyaloshinskii72,Emery82,Nickel06,Giamarchi04} whose bare Hamiltonian for a square lattice of $N_\perp\times N_\perp$ chains of length $L$ is given by 
\begin{widetext}
\begin{equation}
\label{Hamiltonian}
\begin{split} H   =    \sum_{p,\mathbf{k},\sigma}  E_p(\mathbf{k})\, c^\dagger_{p,\mathbf{k},\sigma}c_{p,\mathbf{k},\sigma}   +   {1\over LN^2_\perp} \sum_{\{\mathbf{k},\sigma\}} \,  & \big[\, g_1\, c^\dagger_{+,\mathbf{k}'_1,\sigma_1}c^\dagger_{-,\mathbf{k}'_2,\sigma_2}c_{+,\mathbf{k}_2,\sigma_2}c_{-,\mathbf{k}_1,\sigma_1}  \cr
&    +  \ g_2 \,  c^\dagger_{+,\mathbf{k}'_1,\sigma_1}c^\dagger_{-,\mathbf{k}'_2,\sigma_2}c_{-,\mathbf{k}_2,\sigma_2}c_{+,\mathbf{k}_1,\sigma_1}   \cr
  & + \ g_3 \, \big( c^\dagger_{+,\mathbf{k}'_1,\sigma_1}c^\dagger_{+,\mathbf{k}'_2,\sigma_2}c_{-,\mathbf{k}_2,\sigma_2}c_{-,\mathbf{k}_1,\sigma_1} + \mathrm{H.c} \big)\big]\delta_{\mathbf{k}_1+\mathbf{k}_2= \mathbf{k}'_1+\mathbf{k}'_2(\pm\mathbf{G}) },\end{split}
\end{equation}
\end{widetext}
where the operator $c^\dagger_{p,\mathbf{k},\sigma}$ $(c_{p,\mathbf{k},\sigma}$) creates (destroys) a right ($p=+$) and left $(p=-)$ moving electron of  wave vector  $\mathbf{k}=(k,k_b,k_c)$  and spin $\sigma$.   The free part is  modeled by the  one-electron energy spectrum  
\begin{equation}
\label{spectrum}
\begin{split}
E_p(\mathbf{k} ) =  v_F(pk-k_F)  -2t_{\perp b}\cos k_b  & -2t'_{\perp b}  \cos 2k_b \cr &-2t_{\perp c}\cos k_c,
\end{split}
\end{equation}
where  $v_F$ and $k_F$ are  the longitudinal Fermi velocity and wave vector;    $t_{\perp b}$ and $t_{\perp c}$  are the  nearest-neighbor hopping integrals in the two perpendicular directions. The small transverse second nearest-neighbor hopping   $t_{\perp b}'\ll t_{\perp b}$ paramaterizes the alteration of nesting  of the open  Fermi surface, which simulates the most important effect of pressure in our model.  The   quasi-1D anisotropy  of the spectrum is  \hbox{$E_F\simeq 15 t_{\perp b} \simeq 450 t_{\perp c} $}, where $E_F = v_Fk_F\simeq 3000 $K is the longitudinal Fermi energy congruent with the range  found in the Bechgaard salts\cite{Grant82,Ducasse86,LePevelen01}; $E_F$ is half the bandwidth $E_0=2E_F$ in the model. The interacting part of the Hamiltonian  is described by the bare backward ($g_1$) and forward ($g_2$) scattering amplitudes between right and left moving electrons. In terms of the extended-Hubbard model parameters, $g_1=U-2V$ and $g_2=U+ 2V$, where $U$ and $V$ are the on-site and nearest-neighbor repulsion. The half-filling character of the band -- due to the small dimerization of the chains -- gives rise to  Umklapp scattering of bare amplitude $g_3$ \cite{Emery82},  for which momentum conservation is satisfied modulo  the longitudinal reciprocal lattice vector $\mathbf{G}=(4k_F,0,0)$.

In the repulsive sector, the couplings satisfy  $g_1-2g_2 <g_3 $, a condition that promotes antiferromagnetic  spin fluctuations in the presence of  nesting. In spite of a variety of possibilities for  the  couplings that would be generic for the phenomena we want to discuss,  one can   call  upon experiments and band calculations to delimit their range and make a choice for the amplitude of    normalized couplings $\tilde{g}_i\equiv g_i/\pi v_F$  for the calculations that will follow.  We first note that -- half-filling -- Umklapp term $\tilde{g}_3\approx {\Delta_D\over E_F} \tilde{g}_1$ is proportional to a  small  dimerization gap $\Delta_D$ due to the modulation of the electron transfer integral  along the stacks\cite{Barisic81,Emery82,Penc94}. For (TMTSF)$_2$X compounds, the modulation is relatively small and one finds\cite{Grant82,Ducasse86,LePevelen01} ${\Delta_D\over E_F}\lesssim 0.1$. The backscattering coupling $\tilde{g}_1$ governs spin excitations and is involved in the enhancement of static spin susceptibility   \cite{Bourbon93,Fuseya07}. Experiments in the Bechgaard salts indicate that this enhancement is around 20\%   the non interacting  band value  at low temperature \cite{Miljak85,Miljak83,Wzietek93}. This    is compatible with the  use of a bare backscattering amplitude in the interval   $\tilde{g}_1\simeq 0.3 \ldots 0.5$, giving  in turn the range  $\tilde{g}_3\simeq 0.02 \ldots 0.04$ for Umklapp. As for the forward scattering, its bare amplitude  can be adjusted    in order for the calculated temperature scale for SDW ordering to fall  in the range of observed values 10...20~K    at moderate nesting frustration. This  leads to $\tilde{g}_2 \simeq 0.5 ...0.7$.    Though non exhaustive, this range  of parameters is  found to be generic   of the interdependence between magnetism and superconductivity. The   RG calculations that follow have been carried out  for $\tilde{g}_1= 0.32$, $\tilde{g}_2=0.64$ and $\tilde{g}_3=  0.02$.

 The RG method    consists of integrating successively the degrees  of freedom from the high energy  cut-off   $E_F$ down to the energy   $ {1\over 2}E_0(\ell) = {1\over 2}E_0e^{-\ell}$   above and below the Fermi sheets at step $\ell$.   At the one-loop level, the corrections to the   amplitudes $\tilde{g}_i$ as a function of $\ell$ come from the electron-electron  (Cooper)   and electron-hole (Peierls) scattering channels.    Both  interfere    and generate  momentum dependence for the scattering amplitudes as   $E_0(\ell)=E_0\,e^{-\ell}$ is reduced  with increasing  $\ell$. Here the influence on the  RG flow of the smallest transverse hopping integral, $t_{\perp c}$, is negligible  and has been ignored.  In the three momentum variables scheme of the  renormalization group  adopted here\cite{Nickel06}, each sheet of the Fermi surface  is divided into 32 pieces or patches whose location  defines a particular transverse momentum $k_b$ in   the $b$ direction.  Only the $k_b$ dependence is retained for the couplings, which becomes
$g_i\to g_i(k'_{b1},k'_{b2};k_{b 2},k_{b1})$. The explicit  form of the corresponding flow equations has been given previously [Eqns.(10-12) of Ref.~\cite{Nickel06}]  and  these need not  to be repeated here.

 Following their integration up to $\ell\to\infty$, the presence of a  singularity     in the scattering amplitudes signals  an instability of the normal state at a `critical' temperature\cite{critical} $T_\mu$ . To see   what  kind of order  it refers to,  we  compute the susceptibilities. For the   intrachain interactions given above,  a singularity has been shown  to occur either in the static $\mu\!\!=$SDW or    the $\mu\!\!=$SCd susceptibility $\chi_\mu$\cite{Nickel06,Duprat01}.      In the RG framework, these are expressed as a loop integration     
\begin{equation}
\label{Responses}
\chi_\mu(\mathbf{q},\omega) = {1\over \pi v_F} \int_0^{\tilde{\chi}^0_\mu(\mathbf{q},\omega)}  \langle f_\mu(k_b) z^2 _\mu(k_b)\rangle\,d\ell,
\end{equation}
where $\langle \ldots \rangle $ is an average over $k_b$,   $f_\mu(k_b)=1(\cos k_b)$ is a form factor   for the $\mu=$SDW (SCd) order parameters;  $z_\mu(k_b)$ is the  scaling factor associated to the response function of the channel $\mu$ and which  will be defined shortly.  In  the $T_1^{-1}$ analysis given in Sec.~\ref{SecT1}, the dependence on the   (real) frequency $\omega$ and the three-dimensional wave vector ${\bf q}=(q,q_b,q_c)$  of    $\chi_{\rm {}_{SDW}}$   is needed. It can be introduced  through  the upper bound of the loop  integration, which  will be taken as      the normalized free electron dynamic  susceptibility $\tilde{\chi}^0_\mu  =\pi v_F\chi^0_\mu$ of the SDW channel 
\begin{eqnarray}
 \label{KizeroP}
 \begin{split}
& \tilde{\chi}_{\rm {}_{SDW}}^0 (\mathbf{q}_0 + \mathbf{q},\omega)   = \ln {E_F\over T}  +\psi\Big({1\over 2}\Big)          \\ 
&-{1\over 8\pi^2}\int_{-\pi}^{+\pi}\!\!\!\int_{-\pi}^{+\pi} dk_b dk_c \Big[\psi\Big({1\over 2}Ê+ i{\xi_{\rm P}({\bf k}_\perp,\!\mathbf{q},\omega)\over 4\pi T}\Big) + {\rm c.c}\Big].
 \end{split}
\end{eqnarray}
 $\tilde{\chi}_{\rm {}_{SDW}}^0$ has been  expressed in terms of the deviations $\mathbf{q}$ to the best nesting vector \hbox{$\mathbf{q}_0=(2k_F,\pi,\pi)$}. Here $\psi(x)$ is the digamma function and 
 \begin{equation}
\label{ }
\begin{split}
 \xi_{\rm P}({\bf k}_\perp,\!\mathbf{q},\omega) \simeq  \    v_Fq \, + & \,  (2t_{\perp b}\sin k_b) q_b    -4 t_{\perp b}' \cos 2k_b \cr   +  &(2t_{\perp c}\sin k_c) \sin q_c - \omega .
\end{split}
\end{equation}

As for  $\chi_{{\rm SC}d}$ in the Cooper channel, it will be evaluated in the uniform  \hbox{$ \mathbf{q}=0$} and static \hbox{$\omega=0$} limit, where $ \tilde{\chi}^0_{\rm {}_{SCd}} = \ln E_F/T$. 
 
 Following Ref.\cite{Nickel06}, the flow equation  for the static SDW   vertex part  at $(2k_F,\pi)$
 is given by
 \begin{equation}
\label{ }
\begin{split}
&  \partial_\ell z_{\rm {}_{SDW}}(k_b) =  {1\over 2\pi} \int_{-\pi}^{+\pi}  d\bar{k}_b \,B_P(\bar{k}_b)z_{\rm {}_{SDW}}(\bar{k}_b) \cr & \times [g_2(\bar{k}_b+\pi, k_b,\bar{k}_b,k_b + \pi) + g_3(\bar{k}_b,k_b,\bar{k}_b+\pi,k_b+\pi)],
\end{split}
\end{equation}
which is governed by the combination of couplings $g_2 +g_3$. For the static $\chi_{\rm {}_{SCd}}$,  one has $f_{\rm {}_{SCd}}(k_b)=\cos k_b$ for the form factor and the vertex part at zero pair momentum obeys to 
\begin{equation}
\label{ }
\begin{split}
&\partial_\ell z_{\rm {}_{SCd}}(k_b) = - {1\over 2\pi} \int_{-\pi}^{+\pi}  d\bar{k}_b \,B_C(\bar{k}_b)z_{\rm {}_{SCd}}(\bar{k}_b) \cr & \times [g_1(\bar{k}_b, -\bar{k}_b, {k}_b,-k_b) + g_2(\bar{k}_b, -\bar{k}_b,-k_b, {k}_b)],
\end{split}
\end{equation}
 which  is governed by the combination $-g_1-g_2$. The above expressions depend on the $\ell$-derivative of the Peierls and Cooper loops which read 
\begin{equation} 
\begin{split}
B_{\rm P/\rm C}(\bar{k}_b)& =  
\sum_{\nu=\pm 1} 
\theta \big( \betrag{E_0(\ell)/2  + \nu A_{\rm P/\rm C}(\bar{k}_b) } -E_0(\ell) /2\big) \cr 
 & \times \frac{1}{2} \left( 
\tanh \frac{E_0(\ell)/2+ \nu A_{\rm P/\rm C}(\bar{k}_b)}{2T} 
+ \tanh \frac{E_0(\ell)}{4T} \right) \cr  
&   \ \ \ \ \ \ \ \times \frac{E_0(\ell)/2}{E_0(\ell) +\nu A_{\rm P/\rm C}(\bar{k}_b)} \ , 
\end{split}
\end{equation}
where $A_{\rm P}(\bar{k}_b) = 4 t_{\perp b}'\cos 2k_b$, $A_{\rm C}=0$, and $\theta(x) $ is the step function  with the definition $ \theta(0)\equiv {1\over 2} $.
 \begin{figure}
 \includegraphics[width=6.0cm]{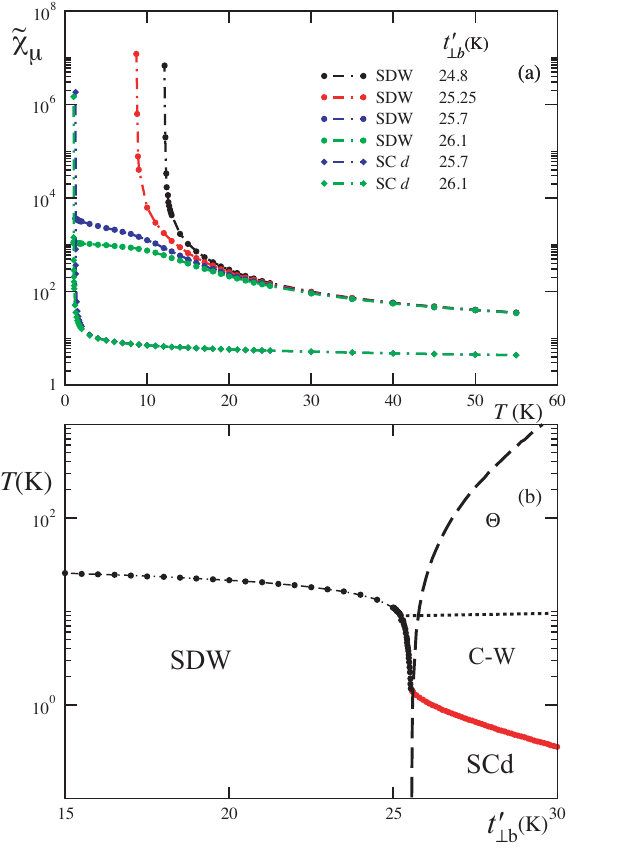}
 \caption{ (a) Temperature variation of the normalized static susceptibility $\tilde{\chi}_\mu$ in the SDW and SCd channels at various $t'_{\perp b}$ on either side of the threshold value $t'^*_{\perp b}$; (b) RG phase diagram of the quasi-1D electron gas model (see text).    The  dashed line stands as the   C-W scale  $\Theta$    in the superconducting  sector. The  dotted line defines  the temperature domain of the C-W  behavior. \label{Ki}} 
 \end{figure}

The RG results at the one-loop level for the temperature dependence of the normalized   $\tilde{\chi}_{\rm {}_{SDW}}$ and $\tilde{\chi}_{\rm {}_{SCd}}$  ($\tilde{\chi}_\mu\equiv \pi v_F \chi_\mu $) are given in Fig.~\ref{Ki} for different values of the nesting frustration parameter $t'_{\perp b}$. At small  $t'_{\perp b}$, $\tilde{\chi}_{\rm {}_{SDW}}$ diverges signaling an instability towards the formation of a SDW state at the temperature $T_{\rm {}_{SDW}}$. This scale decreases as $t'_{\perp b}$ is raised and at the approach of the threshold $t'^*_{\perp b}\simeq 25.6 $K (for the set of parameters used), it undergoes a rapid drop. However, $T_{\rm {}_{SDW}}$ does not go to zero, but is   replaced by another scale $T_c$ at which    $\tilde{\chi}_{\rm {}_{SCd}}$ is singular and  an instability of the metallic state against d-wave superconductivity  takes place. This yields a maximum in $T_c$  when $ T_{\rm {}_{SDW}}$ is minimum, namely where the coupling strength, mediated  by spin fluctuations,  is maximum.   When $t'_{\perp b}$ is further raised, $T_c$     decreases monotonically. 
The overall variation of the temperature scale  for the  metallic state instability  of  Fig.~\ref{Ki}-b captures fairly well the characteristic variation of  the critical temperature found  in compounds like  (TMTSF)$_2$X under pressure\cite{Jerome82,Bourbon08,Brown08}. 
\begin{figure}
 \includegraphics[width=9.0cm]{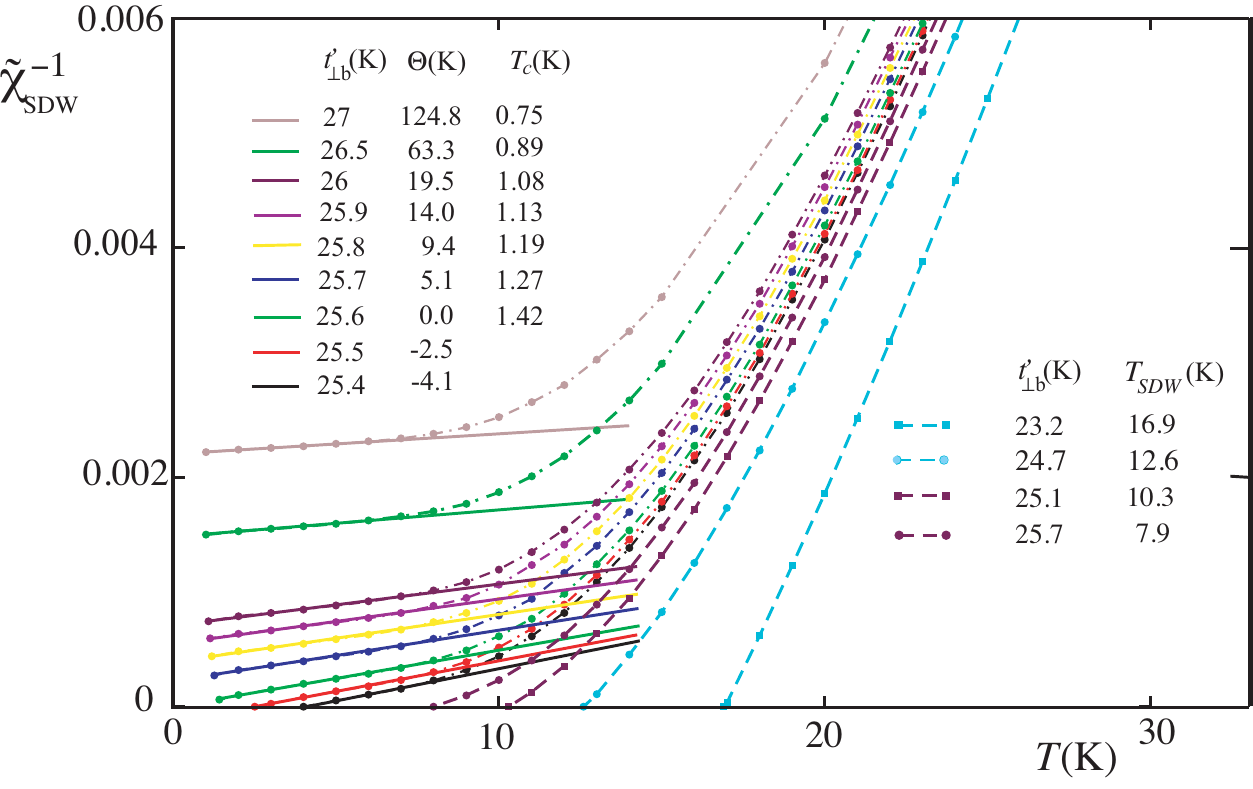} 
  \caption{Calculated inverse normalized SDW susceptibility as a function of temperature for  different $t'_{\perp b}$. The continuous lines correspond to the Curie-Weiss regime.   \label{KiInv}} 
 \end{figure}

The temperature dependence of   \hbox{ $ \tilde{\chi}_{\rm {}_{SDW}}$ } is of particular interest. To begin with the region well below  $  t'^*_{\perp b}$, on the SDW side, namely far from the boundary between SCd and SDW,  the plot of  $\tilde{\chi}^{-1}_{\rm {}_{SDW}}$ in Fig.~\ref{KiInv} shows a linear behavior in temperature down to $T_{\rm {}_{SDW}}$, which is characteristic  of a   $(T-T_{\rm {}_{SDW}})^{-\gamma}$ singularity for    $ \tilde{\chi}_{\rm {}_{SDW}}$ with the classical exponent $\gamma=1$.  As the boundary is approached, however, this behavior for  $\tilde{\chi}^{-1}_{\rm {}_{SDW}}$ does not  extend  down to the critical point, but  evolves towards  another regime  as    deterioration of nesting conditions becomes more perceptible and $T_{{\rm {}_{SDW}}}$ decreases rapidly. By inspection, the latter regime for   $\tilde{\chi}^{-1}_{\rm {}_{SDW}}$ is also found to be linear   sufficiently  close to $T_{{\rm {}_{SDW}}}$,  but with a  smaller slope. The temperature interval above  $T_{{\rm {}_{SDW}}}$ where it takes place, increases  as  $  t'_{\perp b}$ grows.  This regime, which we parameterize by  a C-W form   $\tilde{\chi}^{-1}_{\rm {}_{SDW}} = C (T+\Theta)$, has  a negative scale $\Theta$ that coincides with $T_{\rm {}_{SDW}}$  at $t'_{\perp b}\lesssim  t'^*_{\perp b} $.

As one reaches     $t'_{\perp b}=t'^*_{\perp b} $ where the system is superconducting, $\Theta=0$ and \hbox{$  \tilde{\chi}_{\rm {}_{SDW}} \sim 1/T$}. If superconductivity was absent the system would be   then  quantum critical in the SDW channel  with a singular   antiferromagnetic correlation length $\xi_{a,b}\sim {T}^{-\nu}$, of exponent $\nu=1/2$ in the $ab$ plane. As for the dynamical exponent $z$,  the results of  Sec.~\ref{relax} give    $z=2$. The C-W  form  carries on in the superconducting sector    where $\Theta $  becomes  positive  and grows with  $t'_{\perp b} $. Since $\tilde{\chi}(T\to 0) \to 1/C\Theta$, the  amplitude of $\Theta$ is thus connected to the size of spin fluctuations in the low temperature limit, which decreases rapidly    with  $t'_{\perp b} $. It follows that  $\tilde{\chi}_{\rm {}_{SDW}}$, though no longer singular above $t'^*_{\perp b} $,  is still temperature dependent: despite altered nesting conditions and   scaling towards the formation of a singlet superconducting ground state, antiferromagnetic correlations continue to grow  down to $T_c$. The C-W behavior   extends several    times $T_c$ in the metallic state.

The existence of a C-W behavior above $t'^*_{\perp b} $ is intimately linked to the presence of magnetic  Umklapp scattering ($g_3$) in the model; without this coupling,  $   \tilde{\chi}_{\rm {}_{SDW}} $ is  essentially       flat at low temperature. This is confirmed by putting $g_3=0$ and carrying out the calculation with  the remaining coupling  constants.  Most surprisingly, however,  strongly reduced  nesting of the Fermi surface alone  plays little  role in the enhancement  of Umklapp. The augmentation  turns out to be a consequence of Cooper SCd pairing that reinforces spin fluctuations in the metallic state.  This can be easily checked by setting   $B_C=0$, which removes  all the Cooper pairing terms   from the flow equations of the coupling constants. In  this superconductive pairing-free scheme, which is equivalent to a ladder  diagrammatic (mean-field)  summation in the density-wave pairing  channel alone,  the C-W behavior  is  negligible for     $t'_{\perp b}\ge t'^*_{\perp b} $. 

Interestingly enough,   Umklapp scattering  is coupled to    $g_1$ and $g_2$,  which   both  flow to    strong coupling at the approach of the superconductive fixed point    at  $T_c$. At variance with ordinary s-wave  superconductivity\cite{Umklapp1D}, these scattering amplitudes     are momentum dependent   in a  SCd scenario yielding an  overall positive sign for their  coupling with Umklapp. It follows   that the singular growth of SCd pairing  at low temperature (Fig.~\ref{Ki}-a) interferes  positively with  Umklapp  expanding  the temperature range where  this coupling and in turn  spin fluctuations increase.  It is this  self-consistency between the two pairing channels that is   responsible for   the  C-W law for the staggered  magnetic  susceptibility   down  to $T_c$.    It is worth stressing that the reinforcement of the SDW channel  is not limited to the SC sector, but is also manifest for $t'_{\perp b}\lesssim t'^*_{\perp b}$, where the C-W behavior, with a negative  $\Theta$,  is indicative of  a SDW instability  driven by Cooper pairing.

\section{ Nuclear spin lattice relaxation rate}
\label{relax}
\subsection{Theoretical prediction for the staggered and uniform contributions to nuclear relaxation }
We now turn   to the derivation of the nuclear relaxation rate in the  RG scheme. The $T_1^{-1}$ calculation  starts from the  Moriya expression\cite{Moriya63} 
\begin{equation}
\label{T1Moryia}
T_1^{-1} =T \int |A_{\bf q}|^2\, {\chi''({\bf q},\omega)\over \omega} d^3q,
\end{equation}
which relates $T_1^{-1}$ to the imaginary part of the retarded spin susceptibility $\chi''$. Here $A_{\bf q}$ is proportional to the hyperfine matrix element. The integral over all ${\bf q} $ indicates that $T_1^{-1}$ is sensitive to  staggered and uniform electronic spin correlations with respectively large ${\bf q}\sim {\bf q}_0 $ and small parallel $q\sim0$. We  then consider the following decomposition  
    \begin{eqnarray}
    \label{MoriyaB}
T_1^{-1} & = &T \, \bigg(\int_{q\sim 0}\! +\! \int_{{\bf q} \sim {\bf q}_0}\bigg)  |A_{\bf q}|^2 \,{\chi''({\bf q},\omega)\over \omega} d^3q,\cr\cr
  & \equiv  & T_1^{-1}( {q}\sim 0) + T_1^{-1}({\bf q}_0).   
\end{eqnarray}   
Let us first examine the staggered component, $T_1^{-1}({\bf q}_0) $. Using (\ref{Responses}-\ref{KizeroP}), the expression of $\chi''({\bf q}+ {\bf q}_0,\omega)$ at small ${\bf q}$ and $\omega$ is given by 

\label{SecT1}
\begin{equation}
\label{ }
\begin{split} {\rm Im} \, & \chi_{\rm {}_{SDW}}(\mathbf{q}+\mathbf{q}_0,\omega)  = \cr  &\frac{\chi_{\rm {}_{SDW}}(\mathbf{q}_0)\,\Gamma\,\omega}
	{(1+\xi^2_a\,q^2+\xi^2_b\,q^2_b+\xi^2_c\,(\sin q_c)^2)^2+\Gamma^2\,\omega^2}. \end{split}
\end{equation}
From the results of the \hbox{Appendix~A},  \hbox{$\xi_i^2 = \xi_{0,i}^2 \langle z_{\rm SDW}^2(k_b)\rangle/\tilde{\chi}_{\rm SDW}$}, is the squared of the correlation length along $i=a,b, {\rm and }\  c$ directions;  \hbox{$\xi_{0,a} \propto v_F/T_0 $} and  \hbox{$\xi_{0,b,c} \propto t_{\perp b,c}/T_0 $} are the corresponding coherence lengths evaluated at the SDW temperature $T_0\simeq 12$K obtained at small $t'_{\perp b}$; \hbox{$ \Gamma = \Gamma_0 \langle z_{\rm SDW}^2(k_b)\rangle/\tilde{\chi}_{\rm SDW}$ } is the relaxation time for SDW fluctuations and \hbox{$\Gamma_0 \propto 1/T_0$} is a characteristic    short-range time scale. The integration over ${\bf q}$ is carried out in Appendix A and yields
\begin{equation}
\label{T1q0}
\begin{split}
T_1^{-1}&({\bf q}_0)=  2\pi^3 |A_{{\rm q}_0}|^2 [N(E_F)]^2 v_F\Gamma_0/( \xi_{0,a}\xi_{0,b})  \cr
& \times  T \, \tilde{\chi}_{\rm SDW} \left[ {1\over \sqrt{1+ \xi_c^2}}  - {1\over  \sqrt{(1+r)(1+ r+\xi_c^2)} }\right],\cr
&
\end{split}
\end{equation}
where $r\simeq 1.114\langle z_{\rm SDW}^2(k_b)\rangle/\tilde{\chi}_{\rm SDW}   $ and  $N(E_F) = 1/\pi v_F$ corresponds to  the density of states at the Fermi level. The enhancement of the staggered component as the temperature is lowered is thus  connected to the static SDW response which can be    obtained by the RG method.

We next  consider  the uniform component of the relaxation rate which is connected to the imaginary part of the dynamic spin susceptibility at small $\omega$ and ${ q}$. In this limit, $\chi({\bf q},\omega)$   has been shown  to be non singularly enhanced   by interactions at low temperature \cite{Wzietek93,Bourbon93,Fuseya07}. Within   the random phase approximation, the expression of the imaginary part reads
 \begin{widetext}
\begin{eqnarray}
\label{ImKi0}
\chi''({\bf q}, \omega  )\Big\vert_{\omega, { q} \to 0}\!\! & = &-\eta^2 \!\!{1\over 4\pi^3}\sum_p \int dkdk_b dk_c \, \big[  n\big(E_p({\bf k}+{\bf q})\big)-  n\big(E_p({\bf k})\big) \big] \, \delta\big( \omega- E_p({\bf k}+ {\bf q}) + E_p({\bf k})),
\end{eqnarray}
\end{widetext}
where  $n(x)$ is the Fermi distribution.  The uniform contribution for the imaginary part  is enhanced from  electron-electron interaction by the  factor $\eta$ (see Appendix A). From    previous measurements of the static and uniform spin susceptibility\cite{Miljak85,Miljak83}, its enhancement is about  20\%  in the low temperature domain so that the factor can be fixed to $\eta \simeq 1.2$ considered as    temperature independent in the range of interest \cite{Fuseya05}.  Substituting  in (\ref{MoriyaB}), the  remaining integrals  are carried out in the Appendix A and lead to the `Korringa' component
\begin{equation}
\label{uniform}
T_1^{-1}({ q}\sim 0) =\pi |A_0|^2 [N(E_F)]^2\eta^2 T .
\end{equation}
While the uniform contribution is   nonsingular,  its  amplitude is known to become  ultimately larger than  the staggered component at high enough  temperature \cite{Bourbon93}.
\subsection{\label{Results}Results and relation to  experiments}
\begin{figure}
 \includegraphics[width=9.0cm]{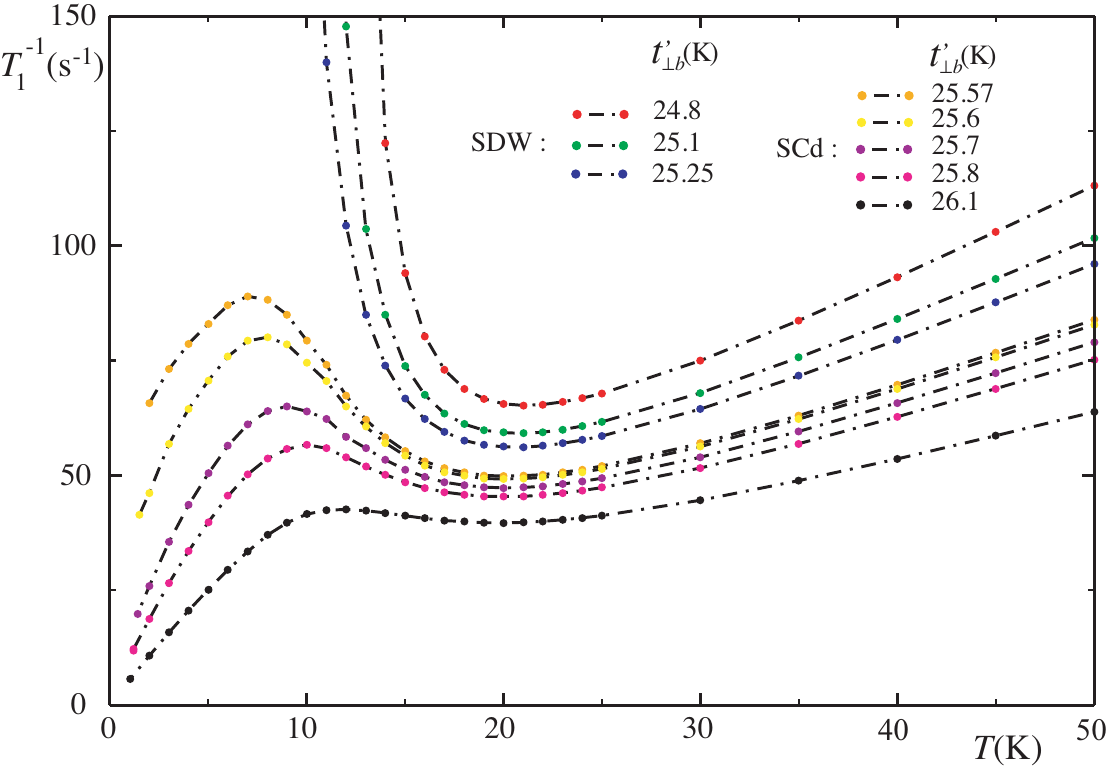}
 \caption{ Calculated temperature profile of nuclear relaxation rate as a function of temperature for $t_{\perp b}'$ below and above the threshold value  $t'^*_{\perp b}$. \label{RateTheory}} 
 \end{figure}

\begin{figure}
 \includegraphics[width=9.0cm]{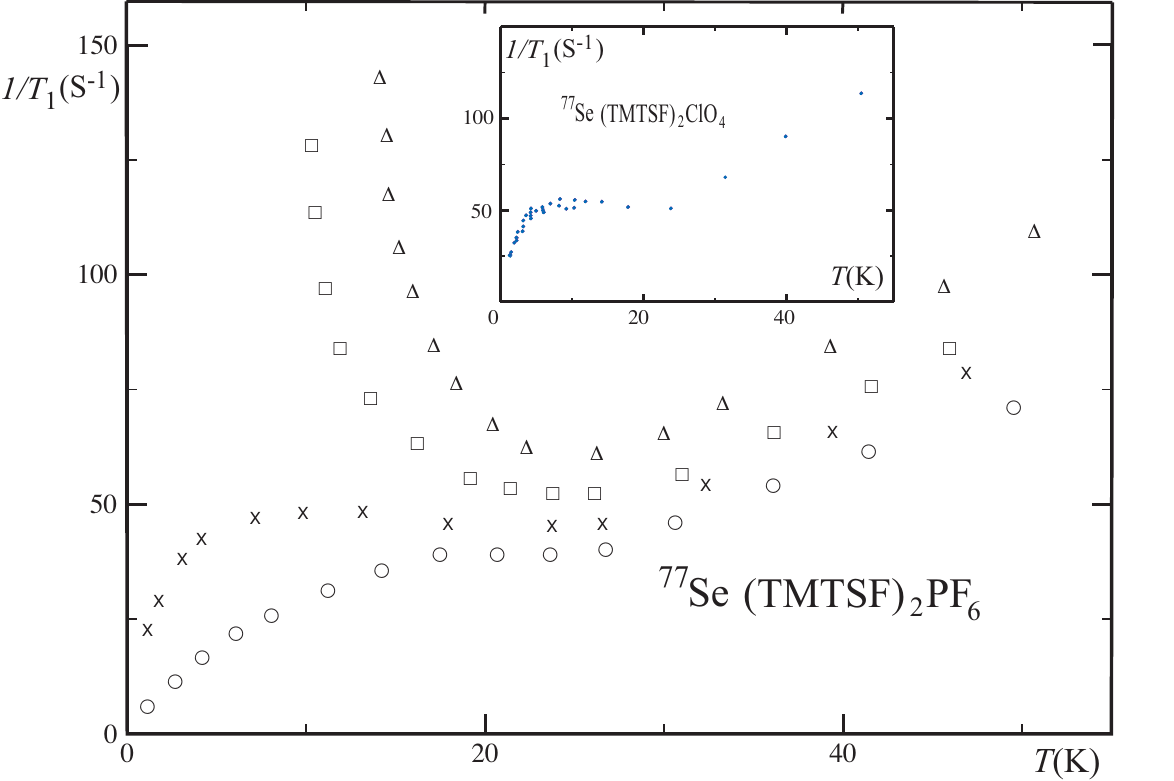}
 \caption{$^{77}$Se $T_1^{-1}$ {\it vs} temperature measured in (TMTSF)$_2$PF$_6$ in the SDW regime at $P=$ 1bar (triangles), 5.5 kbar (squares) and in SC regime at $P=8$kbar (crosses) and  11 kbar (circles) [Ref. \cite{Creuzet87b}]. Inset:  $^{77}$Se $T_1^{-1}$ {\it vs}  $T$ in  (TMTSF)$_2$ClO$_4$ at $P=1$bar [Refs. \cite{Bourbon84,Creuzet85}].\label{PF6ClO4}} 
 \end{figure}

To compare the sum of (\ref{T1q0}) and (\ref{uniform}), as the calculated  $T_1^{-1}$,  with the experimental findings   for  $^{77}$Se $T_1^{-1}$  in    (TMTSF)$_2$PF$_6$ and (TMTSF)$_2$ClO$_4$ [Refs. \cite{Creuzet87b,Bourbon84}],   we   adjust  the two unknown constants $|A_0|$  and  $ |A_{{\bf q}_0}|$ so that  the amplitude  of $T_1^{-1}$  falls in the range of   the observed  values (Fig.~\ref{PF6ClO4}). Since in the  high temperature region,  the nuclear relaxation rate  is  dominated by the uniform contribution \cite{Wzietek93,Bourbon89}, $|A_0|$ is  adjusted  to make  $T_1^{-1}({ q}\sim 0)$ matching with  the measured $T_1^{-1}$ values at 50K.     The other   constant    $ |A_{{\bf q}_0}|$ for the staggered part  is   tuned   such that $T_1^{-1}$ is congruent with  the measured   value at   20K. A ratio of $ |A_{{\bf q}_0}|/|A_0|\sim 10^{-2} $ is thus found  for the hyperfine matrix elements\cite{HFmatrix}.   The total expression of   $T_1^{-1}$ is  then plotted in Fig.~\ref{RateTheory} for various values of the nesting frustration parameter $t'_{\perp b}$, namely below and above the threshold for superconductivity in the calculated phase diagram of Fig.~\ref{Ki}-b.

In the SDW domain for \hbox{$t'_{\perp b} < t'^*_{\perp b}$}, the relaxation rate (\ref{T1q0}) behaves as  $T_1^{-1} \propto T\tilde{\chi}_{{}_{\rm SDW}}/\sqrt{1+ \xi^2_c} $ at large $r$, namely close to $T_{{}_{\rm SDW}} $, where it is dominated by the staggered contribution. The  latter then develops a three-dimensional singularity of the form \hbox{$T^{-1}\sim (T -T_{{}_{\rm SDW}})^{-\upsilon}$}, with $\upsilon=1/2$  when    $\xi_c$ becomes large as  $T\to T_{{}_{\rm SDW}}$.  
The power law exponent   conforms to the one of  mean-field  theory in three dimensions \cite{Bourbon93}. This has been shown long ago   to agree with the  (TMTSF)$_2$PF$_6$ data of Fig.~\ref{PF6ClO4}  \cite{Creuzet87b,Wzietek93}.  As one moves away from $T_{{}_{\rm SDW}}$ in temperature, $\xi_c$ becomes smaller and  the system evolves towards a  two-dimensional   behavior where $T^{-1}\sim T \tilde{\chi}_{{}_{\rm SDW}}$,  corresponding  to $\upsilon =1$. 

Now if one moves along the $t'_{\perp b}$ scale, by approaching $t'^*_{\perp b}$  from below,  one enters into a  transitional regime  where $T_{{}_{\rm SDW}}$ is relatively small; $\xi_c$ then becomes large and three-dimensional order, with $\upsilon =1/2$, develops  only in  very close proximity to the critical point.   For $t'_{\perp b} > t'^*_{\perp b}$  on  the SCd side, $T_1^{-1}$ is no longer singular but shows a pronounced anomaly due to short-range spin  fluctuations. These  extend  deeply in the normal state up to 20 K or so, above which  the uniform component of the relaxation takes over and $T_1^{-1} \propto  T$.  To the anomaly of $T_1^{-1}$  found  down to about 10K (Fig.~\ref{RateTheory}) corresponds a distinct region of increase of $\chi_{{}_{\rm SDW}}$ (Fig.~\ref{Ki}-a) due to the growth of antiferromagnetic correlations close to SDW ordering. $T_1^{-1}$  passes through maximum near 10K, whose amplitude, and to a lesser extend  its location,  are $t'_{\perp b}$ dependent; $T_1^{-1}$ then finally starts to go down at lower temperature  until one reaches $T_c$. 

The overall structure of the calculated $T_1^{-1}$ anomaly compares fairly well with the data of Creuzet {\it et al.,}\cite{Creuzet87b}(Fig.\ref{PF6ClO4})  and Brown {\it et al.,}\cite{Brown08} on (TMTSF)$_2$PF$_6$ above the critical pressure $P_c$ for superconductivity and on (TMTSF)$_2$ClO$_4$ at ambient pressure $(>P_c$, inset of Fig.~\ref{PF6ClO4}). 
 
\begin{figure}
 \includegraphics[width=9.0cm]{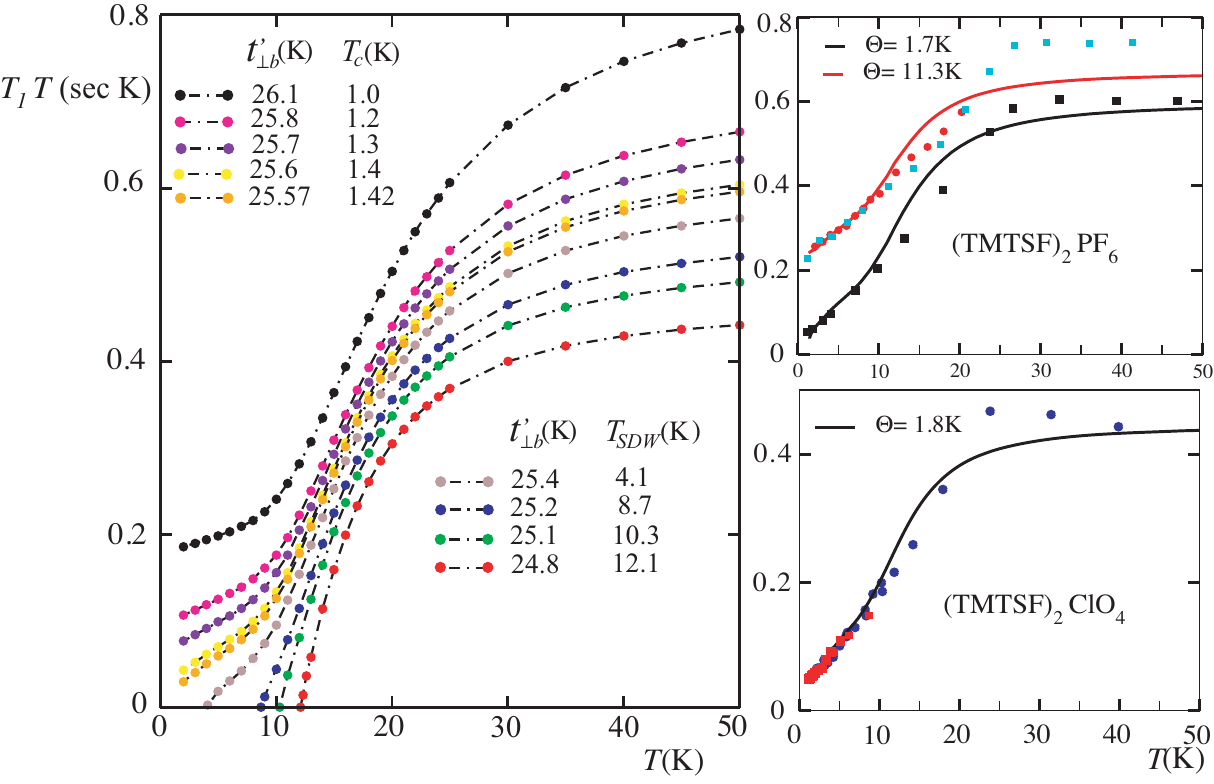}
 \caption{Left: Calculated temperature profile of $T_1T$ for different values of $t_{\perp b}'$ below and above the threshold $ t'^*_{\perp b}$. Upper right: $^{77}$Se $T_1T$ data for (TMTSF)$_2$PF$_6$ at 9.5 kbar (red circles, after Wu {\it al.,} Ref. \cite{Wu05}), 8 kbar and 11 kbar (green squares after Creuzet {\it et al.,} Ref. \cite{Creuzet87b}); the continuous line is a theoretical fit.  Lower right:   $^{77}$Se $T_1T$ data for (TMTSF)$_2$ClO$_4$ at ambient pressure (bleu circles, after Creuzet {\it et al.,} Ref. \cite{Bourbon84}, red circles, after Shinagawa {\it et al.,}  Ref. \cite{Shinagawa07}). \label{T1T}} 
 \end{figure}

If one looks more closely at the decrease of the calculated $T_1^{-1}$  below 10K  in Fig.~\ref{RateTheory}, it is found that it deviates from  a linear Korringa law as a consequence of the growth of spin fluctuations,    responsible   for  the C-W behavior for      $\chi_{{}_{\rm SDW}}$ in this  temperature range  (Fig.~\ref{Ki}). To see  how these  fluctuations mark the relaxation rate, it is instructive  to look at the temperature dependence of  $T_1T$   shown in Fig.~\ref{T1T}. In the left panel of this Figure, the three different regimes of the calculated  relaxation rate can be identified. In the high temperature regime $T_1T$ tends to level off, dominated by the uniform component (\ref{uniform}).  At lower temperature where $T_1T$ is  controlled by the staggered component, two 2D linear regimes,  governed  by  $\xi_{a,b}$, can be singled out and  related   to  those found previously for $1/\tilde{\chi}_{{}_{\rm SDW}}$ in Fig.~\ref{KiInv}. Indeed, in the intermediate temperature range,  between 20 and 10K or so, $T_1T$ is weakly affected by nesting alterations and evolves with a steep slope that would extrapolate to a finite critical temperature.  However, approaching  10K, these alterations become more perceptible and  the slope of $T_1T$ is reduced    and enters in the low temperature  C-W regime  of the form $T_1T = \mathcal{C} (T+ \Theta)$. Following the example of $1/\tilde{\chi}_{{}_{\rm SDW}}$  when  $t'_{\perp b}  \lesssim t'^*_{\perp b}$, $\Theta=-T_{{}_{\rm SDW}}$,  and   $T_1T$ is   found to be linear, except in  the very close vicinity of $T_{{}_{\rm SDW}}$, where  $\xi_c$ becomes  large. At $t'^*_{\perp b}$,  $\Theta$ vanishes    and finally  grows positively for  $t'_{\perp b} > t'^*_{\perp b}$   as  $T_c$ decreases. The slope $\mathcal{C}$ diminishes as $t'_{\perp b}$ grows, whereas  the product $ \mathcal{C} \Theta $, corresponding to the extrapolated zero temperature intercept of $T_1T$,    increases. 

From the NMR work  of Creuzet {\it et al.,}\cite{Bourbon84,Creuzet85} (inset of Fig.~\ref{PF6ClO4}) and from   of Shinagawa {\it et al.,} \cite{Shinagawa07} (Fig.~\ref{T1T}, lower panel on the right) on (TMTSF)$_2$ClO$_4$ at ambient pressure, the three regimes of  $T_1T$ can be discerned  in the data. On the same panel the continuous line corresponds to a fit using the sum of  (\ref{T1q0})  and (\ref{uniform}), where $|A_{{\bf q}_0}|\simeq 18.5~( {\rm sec}^{-1}) $ and  $|A_{{\bf q}_0}|/|A_{0}| \simeq 0.7 \times 10^{-2}$, with a C-W  regime that corresponds to  $\Theta \simeq 1.8$K. In the upper panel of the same Figure, the $T_1T$ data of Wu {\it et al.,} \cite{Wu05} and Creuzet {\it et al.,}\cite{Creuzet87b} (Fig.\ref{PF6ClO4}) for (TMTSF)$_2$PF$_6$ at  $P \simeq 10 $kbar are shown. The fit  (blue curve) is obtained for $|A_{{\bf q}_0}|\simeq12.8 ~({\rm sec}^{-1}) $ and  $|A_{{\bf q}_0}|/|A_{0}| \simeq 0.65 \times 10^{-2}$ for which  $\Theta \simeq 11.3$K. For the data on the same compound at  8 kbar, we have $|A_{{\bf q}_0}|\simeq 18 ~({\rm sec}^{-1}) $ and  $|A_{{\bf q}_0}|/|A_{0}| \simeq 0.9 \times 10^{-2}$, with $\Theta \simeq  1.7$K.  The three temperature regimes are revealed   from the data at small   $\Theta$. However, as pressure increases, in (TMTSF)$_2$PF$_6$, for example, the   temperature interval over which  the C-W  takes place experimentally apparently increases in size. This pressure effect is not captured by the present calculations for  which  the C-W temperature interval is essentially constant as a function of    $t'_{\perp b}$ (Fig. \ref{Ki}-b).  As for the rapid  increase of  $\Theta$ as  the ratio $T_c/T_c( t'^*_{\perp b})$ falls off in Fig. \ref{Ki}-b, it is  found to be   in fair    agreement  with the experimental findings in (TMTSF)$_2$PF$_6$, at least up to moderate pressure where data are available\cite{Brown08,Creuzet87b}. 

\section{\label{SecDisc}Discussion and  Conclusion}
The explanation  put forward for the enhancement of spin fluctuations    in  the metallic state  of  the (TMTSF)$_2$X compounds    modifies an earlier  scheme of interpretation proposed long ago,  in which  an effective -- strongly  renormalized -- scale   for interchain hopping   played the  dominant part in the temperature profile of the enhancement that was considered one-dimensional in character.  In the quasi-one-dimensional  view  adopted  here, the  coherent  wrapping  of the open Fermi surface   takes place  in the  temperature domain  $\sim t_{\perp b}$ ($\sim 100K$), that  is  far above  the range where the anomalous   features of the relaxation rate are found.  It is rather  the small parameter $t'_{\perp b}$  for nesting deviations of the whole Fermi surface that acts as the  critical parameter and triggers the modification of the relaxation rate in temperature. For repulsive couplings,  the increase of $t'_{\perp b}$  alters,  as expected,  the stability of the   SDW fixed point,  simulating   the  effect of pressure\cite{Horovitz75,Yamaji82}. At  some threshold value $t'^*_{\perp b}$, the SDW  fixed point  is    unstable. However, because  of a finite mixing between the weakened density-wave and  unaltered Cooper pairing singularities in the scattering amplitudes, the electron system is not  a Fermi liquid, but is rather  characterized  by  superconducting order, which   takes place in the SCd channel for intrachain repulsive interactions. 

We have first seen how  this crossover between  fixed points operates  as  a function of temperature above $t'^*_{\perp b}$ for the SDW response function.  At high  temperature,    thermal broadening of the Fermi surface makes nesting deviations  less perceptible and  the  electron system  is still    attracted  by the   SDW primary fixed point.   As   the temperature decreases  and the fine details of the Fermi surface in the $ab$ plane  become  progressively  coherent, the singularity  of the SDW  response is  suppressed. This coincides with the  emergence  of a  secondary   SCd fixed point, whose influence stretches in temperature about ten times the maximum $T_c$  value reached at $t'^*_{\perp b}$.  Throughout the flow toward $T_c$, SDW correlations, albeit non singular,   persist  to increase in the $ab$ plane, thanks to the strengthening of  Umklapp    by Cooper SCd pairing.  The increase of  the SDW susceptibility  can be fitted  with a Curie-Weiss law in temperature.

 The above features   found for the susceptibility are also encountered in the antiferromagnetic  component of   the nuclear relaxation rate as an anomalous enhancement  that emerges  out of  a `Korringa' or Fermi liquid like behavior      at low  temperature. These characteristics of the nuclear  relaxation rate  adhere to a large extent to the experimental facts  found by NMR in the Bechgaard salts. The  relatively rapid evolution of the relaxation rate enhancement seen  under pressure, in particular concerning the characteristics of the Curie-Weiss law, can find an explanation in  line with the strong reduction of $T_c$ under pressure. This connection between and experiment gives significant  support  to  a mechanism of superconductive pairing     mediated by spin fluctuations.

 As stressed before, at the core of  the weak coupling scaling  theory resides the  finite quantum interference  between   density-wave and Cooper pairing channels. It  gives rise to d-wave superconductivity   from the exchange of spin correlations, and conversely to the enhancement of spin correlations from  superconductive pairing. This interference is manifest in   the one-loop perturbation theory. Higher order  effects  like the   interaction between SDW modes of fluctuations are neglected. In self-consistent renormalized theory of spin fluctuations where such  mode-mode interactions are included  \cite{Moriya90},  a Curie-Weiss enhancement  can be found in the high temperature part of the normal phase. In such an approach,  the SDW   channel is singled out and  interference with Cooper pairing absent.  It follows that  C-W enhancement does not persist  down to the lowest temperature, where instead a Fermi liquid behavior takes place.  Despite this fundamental difference,  one cannot exclude that  the observed   Curie-Weiss enhancement of the nuclear relaxation rate    superimposes to some extent both contributions.

In conclusion the  above results have highlighted  that  from a  renormalization group approach to  the quasi-one-dimensional electron gas model, it is possible to obtain a  microscopic  description of  the spin fluctuations,   as extracted from nuclear relaxation in (TMTSF)$_2$X. It is through the same approach that   the  mechanism of   interplay  between    itinerant antiferromagnetism and superconductivity   has been   worked out for the    phase diagram of these compounds\cite{Nickel06,Duprat01}. In parallel with the  work presented here for the relaxation rate, the same approach  has been applied to  electron transport in the Bechgaard salts, which also shows an anomalous temperature dependence in the metallic state above superconductivity. Calculations of the transport scattering rate are compared to the resistivity in a separate paper \cite{DoironLeyraud09}.


\begin{acknowledgments}
We thank P. Auban-Senzier, N. Doiron-Leyraud, D. J\'erome, L. Taillefer and A.-M. Tremblay  for useful comments on several aspects of this work.  C.~B.   expresses his gratitude to  the Natural
Sciences and
Engineering Research Council of Canada (NSERC),  and the
 Canadian Institute for Advanced  Research   (CIFAR)  for
financial support. The authors are  thankful to the R\'eseau Qu\'eb\'ecois de Calcul Haute Performance (RQCHP) for supercomputer facilities at the Universit\'e de Sherbrooke. 

\end{acknowledgments}
\medskip

\begin{appendix} 
\section{Nuclear relaxation rate}
\subsection{Antiferromagnetic part}
The wave vector and frequency dependence of the dynamic susceptibility is in general not given by the RG method used here. However, an expression for the  imaginary part of the SDW susceptibility can be obtained by  restoring the $q$ and $\omega$ dependence through the boundary conditions of the flow equation for the susceptibility   (\ref{Responses}). Using (\ref{KizeroP}) at small $\mathbf{q}$ and $\omega$, one finds 
\begin{equation}
\label{ImKi}
\begin{split}
{\rm Im} \, & \chi_{\rm {}_{SDW}}(\mathbf{q}+\mathbf{q}_0,\omega) =  {1\over \pi v_F}{\rm Im}\!\! \left[\int_0^{\tilde{\chi}^0_{\rm {}_{SDW}}(\mathbf{q}+\mathbf{q}_0 ,\omega)}  
\!\! \!\! \langle z^2 _{\rm {}_{SDW}}(k_b)\rangle\,d\ell\right] \cr\cr
 = &\,\chi_{\rm {}_{SDW}}(\mathbf{q}_0)
	  {\rm Im} \bigr[1-\xi^2_a q^2-\xi^2_b q^2_b-\xi^2_c (\sin q_c)^2+i\Gamma\omega+\dots \bigl]\end{split}
	\end{equation}
	This expression will be  equated with the expansion   of 
\begin{equation}
\begin{split}	
{\rm Im} \, & \chi_{\rm {}_{SDW}}(\mathbf{q}+\mathbf{q}_0,\omega) \cr & \simeq {\rm Im}\,\Bigr[\frac{\chi_{\rm {}_{SDW}}(\mathbf{q}_0)}
	{1+\xi^2_a\,q^2+\xi^2_b\,q^2_b+\xi^2_c\,(\sin q_c)^2-i\Gamma\omega}\Bigl]\cr\cr
&= \frac{\chi_{\rm {}_{SDW}}(\mathbf{q}_0)\,\Gamma\,\omega}
	{(1+\xi^2_a\,q^2+\xi^2_b\,q^2_b+\xi^2_c\,(\sin q_c)^2)^2+\Gamma^2\,\omega^2},\cr
\end{split}
\end{equation}
which we shall use in the following.  Here \hbox{$\xi_i^2 = \xi_{0,i}^2 \langle z_{\rm {}_{SDW}}^2(k_b)\rangle/\tilde{\chi}_{\rm {}_{SDW}}$}, is the square  of correlation length along $i=a,b, {\rm and }\  c$ directions; \hbox{$\xi_{0,i}$} are the corresponding coherence lengths evaluated at the SDW 
temperature $T_0\simeq 12$K obtained for small $t'_{\perp b}$, that is 
 \begin{equation}
\xi_{0,i}^2 = - \! \left({v_i \over 2\pi T_0}\right)^{\! 2}\!\! {\rm Re}\!\left[\int_0^{\pi/ 2}\!{dk_b\over 2\pi} 
	\psi^{''} \!\Big( {1\over 2}-i\,{t'_{\perp b}\over \pi T_0}\cos(2k_b)\Big) \right],\end{equation}
	where $v_a=  v_F/\sqrt{2 }$ and  $v_{b,c}= t_{\perp b,c}$. The relaxation time for  SDW  fluctuations is \hbox{$ \Gamma = \Gamma_0 \langle z_{{}_{\rm {}_{SDW}}}^2(k_b)\rangle/\tilde{\chi}_{{}_{\rm {}_{SDW}}}$}  where 
\hbox{$\Gamma_0 $} is a characteristic short-range time scale at short distance, which  is given by:
\begin{equation}
\Gamma_0={1 \over \pi T_0} {\rm Re }\left[\int_0^{\pi/ 2}{dk_b\over 2\pi} 
	\psi^{'} \Big( {1\over 2}-i\,{t'_{\perp b}\over \pi T_0}\cos(2k_b)\Big) \right].
\end{equation}

Substituting the imaginary part (\ref{ImKi}) in the expression for  the  antiferromagnetic component of the  nuclear relaxation rate (\ref{MoriyaB}),   we get in the limit  \hbox{$\omega \rightarrow 0$}:
\begin{widetext}
\begin{eqnarray}
 T_1^{-1}(\mathbf q \sim  \mathbf q_0)  &=& 8 T |A_{\mathbf{q}_0}|^2\displaystyle
	\int_0^{\xi_{0,a}^{-1}}\int_0^{\xi_{0,b}^{-1}} \int_0^{\pi}  {{\rm Im}\chi({\mathbf q},\omega)\over \omega} \,{dq dq_b dq_c 
}\cr\cr
   & = & 8\pi|A_{\mathbf q_0}|^2  T [N(E_F)]^2 \displaystyle \tilde{\chi}_{\rm {}_{SDW}}(\mathbf q_0){v_F \Gamma_0 \over \xi_{0,a}\xi_{0,b}} 
	 \int_0^{\sqrt{r\over \alpha}}\int_0^{\sqrt{r\over \alpha}}\int_0^{\pi} \frac{dx dy  dq_c}{(1+x^2+y^2 +\xi^2_c\,(\sin q_c)^2)^2}\cr\cr
	 &=& { 2\pi^3}|A_{\mathbf q_0}|^2  T [ N(E_F)]^2 \displaystyle \tilde{\chi}_{\rm {}_{SDW}}(\mathbf q_0){v_F \Gamma_0 \over \xi_{0,a}\xi_{0,b}} \left[ {1\over \sqrt{1+ \xi_c^2}}  - {1\over  \sqrt{(1+r)(1+ r+\xi_c^2)} }\right], 
\end{eqnarray}
\end{widetext}
Here  $r=  \alpha{\langle z^2 _{\rm {}_{SDW}}(k_b)\rangle/\tilde{\chi}_{\rm {}_{SDW}}}$ and $\alpha\simeq 1.114$ is a   constant introduced to adjust the upper bound cut-off in order to go through  a  polar integration in the {\it ab}  plane. This allows   an  analytical expression to be found  that deviates from the numerical (rectangular) integration   by less than 0.1\%. 
\subsection{Uniform part}
Following Eq.~(\ref{MoriyaB}), the uniform component of the relaxation rate is given by 
\begin{equation}
\label{T0}
T_1^{-1}( q \sim  0) =T |A_{0}|^2\int_{ q \sim 0} \, {{\rm Im}\chi({\mathbf q},\omega)\over \omega} {d^3q\over (2\pi)^3},
\end{equation}
Using an RPA expression for the dynamical spin susceptibility at small ${ q}$ and $\omega$, we have for the imaginary part
\begin{equation}
\begin{split}
& {\rm Im}\chi({\bf q}, \omega  )\cr  & = -\eta^2{2\over L\,N_\perp}\sum_{\mathbf k}\sum_p\, {\rm Im}\Big[  
	{n\big(E_p({\bf k}+{\bf q})\big)-  n\big(E_p({\bf k})\big) \over E_p({\bf k}+{\bf q})-E_p({\bf k})-\omega-i0^+}\Big] \cr
	 \cr
		\end{split}
	\end{equation}
where $\eta = (1-\lambda)^{-1}$ is an  enhancement factor from interaction parameterized by $\lambda <1$. For $\omega$ and ${ q}$ going to zero, we have in the low temperature limit 	
\begin{widetext}
\begin{equation}
\begin{split}
{\rm Im}\chi({\bf q}, \omega  )\Big\vert_{\omega, { q} \to 0}
& = - \eta^2{1\over 4\pi^2}\sum_p \iint dE_p{dS_{E}\over | \nabla E_p ({\bf k})|  }\, 
	\big[  n\big(E_p({\bf k})+\omega\big)-  n\big(E_p({\bf k})\big) \big] 
	\delta\big( \omega- E_p({\bf k}+ {\bf q}) + E_p({\bf k})),\\
&= \eta^2 {\omega\over 4\pi^2}  \sum_p \int {dS_{\rm F}\over | \nabla E_p ({\bf k}_{\rm F})|  }\,
	\delta\big( E_p({\bf k}_{\rm F} + {\bf q}) - E_p({\bf k}_{\rm F})). 	\end{split}
\end{equation}
\end{widetext}
Only a constant (Fermi) surface integral remains.  The substitution in  \eqref{T0}, allows one to write 
\begin{widetext}
\begin{equation}
\begin{split}
T_1^{-1}( q \sim  0) =&{T \over (2\pi)^5}|A_{0}|^2\eta^2\sum_p \int d^3q\int {dS_{\rm F}\over | \nabla E_p ({\bf k}_{\rm F})|}
	\delta\big( E_p({\bf k}_{\rm F} + {\bf q}) - E_p({\bf k}_{\rm F}))\cr =&{2T\over (2\pi)^5} |A_{0}|^2\eta^2 \Big(\int {dS_{\rm F}\over | \nabla E_p ({\bf k}_{\rm F})|}\Big)^2
\end{split}
\end{equation}
\end{widetext}
Using the approximation $| \nabla E_p ({\bf k}_{\rm F})|\approx v_F$  for $t_{\perp c}\ll t_{\perp b}\ll v_F$, and $\int dS_{\rm F}=4\pi^2$, one obtains 
 the `Korringa' component
\begin{equation}
\label{ }
T_1^{-1}({q}\sim 0) =  \pi |A_0|^2 [N(E_F)]^2\eta^2T.
\end{equation}
\end{appendix}
\bibliography{articlesII}

\begin{thebibliography}{40}
\bibitem[{cri()}]{critical}
\bibinfo{note}{For anisotropic two-dimensional electron systems, $T_\mu$ is a
  scale for the onset of short-range order in the channel $\mu$. True
  long-range order, at a temperature nevertheless close to $T_\mu$, is assured
  by the finite interchain hopping $t_{\perp c}$ along the $c$ direction.}
\end{thebibliography}

\end{document}